# Entanglement measures for multipartite quantum systems: limitations and new approaches

[1,a] Reza Hamzehofi

[1] Department of Physics, Faculty of Science, Shahid Chamran University of Ahvaz, Ahvaz, Iran

**Abstract.** In this research, we investigate the distribution of entanglement within two entangled $n$-qubit systems using the one-tangle and $\pi$-tangle. Our analysis reveals that for certain quantum states such as the generalized $W$ state, where the probability coefficients depend on the number of qubits, increasing the system size causes these traditional measures to gradually diminish, with the monogamy relation tending toward equality. This behavior highlights a limitation of conventional tangle-based measures in capturing multipartite correlations in large systems. To overcome this, we introduce two alternative measures: the sum of squared one-tangles and the generalized residual entanglement. Unlike the standard one-tangle and $\pi$-tangle, these new quantities remain robust as the number of qubits increases. Moreover, we propose a strong monogamy relation that preserves nontrivial entanglement distribution even in the large-$n$ limit. Our results provide a refined framework for quantifying entanglement in high-dimensional multipartite systems.

## 1 Introduction

Quantum entanglement is a striking phenomenon that defies classical notions of locality and independence, establishing deep correlations between particles regardless of distance. This intricate connection between quantum systems has profound implications for our understanding of reality and the foundations of physics. The study of entanglement in many-body quantum systems is crucial, as it provides deeper insights into the complex behaviors of large quantum systems. Understanding these entanglement properties is essential for unlocking the potential of quantum technologies, including quantum computing, communication, and cryptography, where entanglement plays a key role in achieving breakthroughs in processing power and secure data transmission [1-10]. Quantifying entanglement is crucial for understanding the complexity of quantum systems. As the number of qubits grows, analyzing the entanglement structure becomes vital for evaluating quantum states. By measuring entanglement, researchers can assess the computational capabilities and potential limitations of quantum algorithms. This also enables the development of effective communication protocols, quantum error correction methods, and a deeper understanding of the principles underlying quantum information processing [11-15].

To evaluate the entanglement of pure and mixed $n$-qubit systems, established measures such as one-tangle, $\pi$-tangle, and negativity have been widely used [16-22]. However, one of the fundamental questions about these measures is whether they remain effective and reliable for characterizing the entanglement of high-dimensional systems. As the dimensionality of quantum systems increases, entanglement properties become more intricate, requiring measures that can accurately capture complex correlations. Therefore, it is crucial to investigate the scalability, applicability, and limitations of these entanglement measures when applied to larger quantum systems. The structure of this paper is as follows. First, a brief description of the monogamy of entanglement and the measures used in this study will be provided. We also prove that the sum of squared one-tangles serves as a valid entanglement measure for $n$-partite systems. Next, in Section 3, we analyze specific quantum states, such as the $W$ state. We observe as the number of particles increases, the one-tangle and $\pi$-tangle gradually decrease, showing limited sensitivity in large systems.

[a] e-mail: rezahamzehofi@gmail.com (corresponding author)

Additionally, it is shown that monogamy of entanglement tends toward equality in these cases. Finally, the conclusion is presented in Section 4.

## 2 Eetanglement measures of *n*-qubit systems

### 2.1 The monogamy of entanglement in terms of negativity

For a bipartite system $\rho_{AB}$, the negativity is given by [20]:

$$N_{AB} = 2\sum_{\lambda_i<0}|\lambda_i| \tag{1}$$

where $\lambda_i$ are the negative eigenvalues of the partial transpose $\rho_{AB}^{T_\alpha}$, with respect to $\alpha \in \{A,B\}$. Negativity serves as a widely used entanglement measure for bipartite systems, effectively capturing the degree of entanglement between two subsystems. When extending this concept to a tripartite system $\rho_{ABC}$, we encounter additional complexity in the distribution of entanglement across the three subsystems. Specifically, the entanglement between subsystem *A* and subsystem *B* (measured by the negativity $N_{AB}$) and the entanglement between subsystem *A* and subsystem *C* (measured by the negativity $N_{AC}$) cannot be maximized simultaneously. This interplay is governed by the monogamy of entanglement, which restricts how entanglement can be shared among multiple parties. The monogamy inequality is expressed as [23]:

$$N_{A(BC)} \geq N_{AB} + N_{AC} \tag{2}$$

where, $N_{AB}$ and $N_{AC}$ are referred to as the two-tangle for their respective pairs of subsystems. Moreover, the one-tangle $N_{A(BC)}$ is the negativity of subsystem *A* with respect to the joint system *BC*.

For an multipartite state $\rho_{123...n}$, the monogamy of entanglement is characterized as follows [24, 25]:

$$N_{i|\bar{i}}^2 \geq \sum_{k=1}^{n} N_{ik}^2 \tag{3}$$

where the bipartition between subsystem *i* and all the remaining subsystems is denoted by $N_{i|\bar{i}}$ and $N_{ik}$ denotes the two-tangle with respect to the *i*-th and *k*-th parties. Additionally, the one-tangle is defined as follows [20]:

$$N_{i|\bar{i}} = 2\sum_{\chi_i<0}|\chi_i| \tag{4}$$

where $\chi_i$ are the negative eigenvalues of the partial transpose $\rho_{123...n}^{T_i}$.

The residual entanglement, denoted by $\pi_i\left(i=1,2,...,n\right)$, quantifies the remaining entanglement associated with each subsystem and is defined as follows:

$$\pi_i = N_{i|\bar{i}}^2 - \sum_{k=1}^{n} N_{ik}^2 \tag{5}$$

In an asymmetric system, we have: $\pi_1 \neq \pi_2 \neq ... \neq \pi_n$. Then, the $\pi$-tangle which quantify how much $\rho_{123...n}$ is entangled is defined as follows:

$$\pi = \frac{1}{n}\sum_{i=1}^{n}\pi_i \tag{6}$$

In a symmetric system, since $\pi_1 = \pi_2 = ... = \pi_n$, the $\pi$-tangle can be calculated with respect to an arbitrary qubit using Eq. (5).

### 2.2 Sum of squared one-tangles

For an multipartite state $\rho_{123...n}$, this study suggests analyzing the entanglement by summing the squared one-tangles.

$$Sum\left(N_{i|\bar{i}}^2\right) \equiv \sum_{i=1}^{n} N_{i|\bar{i}}^2 \tag{7}$$

Now we prove that the above relation constitutes an entanglement measure for $n$-partite systems.

A valid measure of entanglement must satisfy the following conditions [20]:

a) Zero for separable states: The measure is zero for separable states, indicating no entanglement.
b) Monotonicity: The measure doesn't increase under local operations and classical communication (LOCC).
c) Invariance under local unitary operations: The measure remains unchanged by local unitary transformations.

For an arbitrary pure state $\rho_{123\ldots n}$, if the sum of squared one-tangles is zero, each individual one-tangle must also be zero. If $N_{i|\bar{i}}=0$, then we have: $\rho_{i|\bar{i}}=\rho_i\otimes\rho_{\bar{i}}$. Therefore, if all the one-tangles vanish, the system is separable. In this way, the first condition is proven.

To prove condition (b), we notice that the one-tangle is an entanglement monotone under local operations and classical communication (LOCC). Specifically, for any LOCC operation $\Lambda_{LOCC}$, we have: $N_{i|\bar{i}}\left(\Lambda_{LOCC}\right)\le N_{i|\bar{i}}\left(\rho_{123\ldots n}\right)$. Since one-tangle is monotonic under LOCC, the squared one-tangle also satisfies the same inequality. Summing over all subsystems it follows that $\sum_i N_{i|\bar{i}}^2\left(\Lambda_{LOCC}\right)\le\sum_i N_{i|\bar{i}}^2\left(\rho_{123\ldots n}\right)$. Then the sum of squared one-tangles doesn't increase under local operations and classical communication (LOCC).

To prove condition (c), consider a local unitary operation of the form $U=U_1\otimes U_2\otimes\cdots\otimes U_n$, acting on the entire system. The transformed state is $\rho'_{123\ldots n}=U\rho_{123\ldots n}U$. The one-tangle is invariant under local unitaries, i.e., $N_{i|\bar{i}}\left(\rho'_{123\ldots n}\right)=N_{i|\bar{i}}\left(\rho_{123\ldots n}\right)$. Consequently the squared one-tangle satisfies this relation too. Summing over all subsystems, we obtain $\sum_i N_{i|\bar{i}}^2\left(\rho'_{123\ldots n}\right)=\sum_i N_{i|\bar{i}}^2\left(\rho_{123\ldots n}\right)$. Hence, the sum of squared one-tangles is invariant under local unitary operations.

## 3 Results

In this section, the entanglement of the generalized $W$ state and the multi-qubit state $|\xi\rangle=1/\sqrt{n+1}\left(|1\rangle^{\otimes n}+|100\ldots0\rangle+\ldots+|000\ldots1\rangle\right)$ are investigated. Both states have probability coefficients that explicitly depend on the number of qubits. Therefore, the one-tangle, π-tangle, and the monogamy relations are also expected to depend on the system size. By analyzing these states, we aim to study how multipartite entanglement measures scale with the number of qubits. In particular, this allows a direct comparison between the sum of squared one-tangles and the conventional one-tangle and π-tangle in multi-qubit systems.

### 3.1 The generalized *W* state

To obtain the $\pi$-tangle of the generalized $W$ state, first, it is necessary to calculate the partial transpose of the system and the partial transposes of the bipartitions. Since the generalized $W$ state is symmetric, the entanglement can be obtained through Eq. (5) with respect to an arbitrary qubit. For this purpose, we consider the first qubit. The density matrix of the generalized $W$ state is given as follows:

$$\begin{aligned}\rho_W=\frac{1}{n}&\\ &\left(|100\ldots0\rangle\langle100\ldots0|+\ldots+|100\ldots0\rangle\langle000\ldots1|\right.\\ &+|010\ldots0\rangle\langle100\ldots0|+\ldots+|010\ldots0\rangle\langle000\ldots1|\\ &\vdots\\ &\left.+|000\ldots1\rangle\langle100\ldots0|+\ldots+|000\ldots1\rangle\langle000\ldots1|\right)\end{aligned}\tag{8}$$

The partial transposes $\rho_W^{T_1}$ can be obtained simply by swapping the first qubits:

$$\rho_W^{T_1} = \frac{1}{n} \begin{aligned} &\big(|100...0\rangle\langle 100...0| + \ldots + |000...0\rangle\langle 100...1| \\ &+ |110...0\rangle\langle 000...0| + \ldots + |010...0\rangle\langle 000...1| \\ &\vdots \\ &+ |100...1\rangle\langle 000...0| + \ldots + |000...1\rangle\langle 000...1|\big) \end{aligned} \tag{9}$$

Since only the matrix elements associated with the following bases of the Hilbert space are non-zero:

$$\begin{aligned} &\{|000...0\rangle, |100...0\rangle, |010...0\rangle, ..., |000...1\rangle, \\ &|110...0\rangle, |101...0\rangle, ..., |100...1\rangle\} \end{aligned} \tag{10}$$

the matrix form of $\rho_W^{T_1}$ in terms of the bases (10) can be written as follows:

$$\rho_W^{T_1} = \frac{1}{n} \begin{pmatrix} 0 & 0 & 0 & \cdots & 0 & 1 & 1 & 1 & \cdots & 1 \\ 0 & 1 & 0 & \cdots & 0 & 0 & 0 & 0 & \cdots & 0 \\ 0 & 0 & 1 & \cdots & 1 & 0 & 0 & 0 & \cdots & 0 \\ \vdots & \vdots & \vdots & \ddots & \vdots & 0 & 0 & 0 & \cdots & 0 \\ 0 & 0 & 1 & \cdots & 1 & 0 & 0 & 0 & \cdots & 0 \\ 1 & 0 & 0 & \cdots & 0 & 0 & 0 & 0 & \cdots & 0 \\ 1 & 0 & 0 & \cdots & 0 & 0 & 0 & 0 & \cdots & 0 \\ 1 & 0 & 0 & \cdots & 0 & 0 & 0 & 0 & \cdots & 0 \\ \vdots & \vdots & \vdots & \cdots & \vdots & \vdots & \vdots & \vdots & \cdots & 0 \\ 1 & 0 & 0 & \cdots & 0 & 0 & 0 & 0 & \cdots & 0 \end{pmatrix} \tag{11}$$

where the arrangement of the matrix bases corresponds to the ordering of the reduced Hilbert space given in Eq. (10).

The non-zero eigenvalues of $\rho_W^{T_1}$ are obtained as follows:

$$\begin{aligned} &\lambda_1 = \frac{n-1}{n}, \\ &\lambda_{2,3} = \pm\frac{\sqrt{n-1}}{n}, \\ &\lambda_4 = \frac{1}{n} \end{aligned} \tag{12}$$

Since the only negative eigenvalue of $\rho_W^{T_1}$ is $-\sqrt{n-1}/n$, using Eqaution (4), the one-tangles are obtained as follows:

$$N_{i|\bar{i}}(W) = \frac{2\sqrt{n-1}}{n} \tag{13}$$

It is found that $\underset{n\to\infty}{Limit}\, N_{i|\bar{i}}(W) = 0$; As the number of qubits increases, the contribution of entanglement between the $i$-th qubit and the rest of the system becomes smaller because the entanglement is being spread across a larger number of qubits. However, the system as a whole remains maximally entangled in the sense that the entanglement is still evenly distributed across all qubits.

Next, to obtain the two-tangles, the reduced density matrices of the $i$-th and $k$-th parties $(\rho_{ik})$ are required. By performing $n-2$ successive partial traces over the density matrix (8), the reduced density matrices $\rho_{ik}$ are obtained as follows:

$$\rho_{ik} = \frac{1}{n} \begin{pmatrix} n-2 & 0 & 0 & 0 \\ 0 & 1 & 1 & 0 \\ 0 & 1 & 1 & 0 \\ 0 & 0 & 0 & 0 \end{pmatrix} \tag{14}$$

Now, after obtaining the partial transpose $\rho_{ik}^{T_i}$ (or $\rho_{ik}^{T_k}$ ) and using Eq. (1), the two-tangles are calculated as follows:

$$N_{ik}(W) = \frac{\sqrt{(n-2)^2+4} - n + 2}{n} \tag{15}$$

Using Eqs. (5), (13) and (15), the $\pi$-tangle is obtained as follows:

$$\begin{aligned} \pi_W = &\left(\frac{2\sqrt{n-1}}{n}\right)^2 - (n-1) \\ &\times \left(\frac{\sqrt{(n-2)^2+4} - n + 2}{n}\right)^2 \end{aligned} \tag{16}$$

In a $W$ state the entanglement is highly delocalized, meaning that each qubit is only weakly entangled with the others, but the system as a whole remains robustly entangled.

However, based on Eq. (16), since $\pi_W$ is a decreasing function of the number of qubits, it tends to zero as the number of qubits becomes large, it does not fully capture this delocalized form of entanglement. This measure is more sensitive to states where the probability coefficients are not dependent on the number of qubits, such as GHZ state.

Fig. 1 shows the one-tangle (Eq. (13)), two-tangle (Eq. (15)), and $\pi$-tangle (Eq. (16)) of a generalized $W$ state as a function of the number of qubits. The decrease of these three measures with the increase in the number of qubits causes the monogamy of entanglement given in relation (3) to approach equality. One important result is that while the one-tangle and $\pi$-tangle adequately quantify the entanglement of the $W$ state for a small number of qubits, their values decrease significantly as the number of qubits increases. Consequently, these measures may misleadingly suggest that the system is nearly separable, even though the $W$ state remains genuinely entangled. This behavior occurs because the probability amplitudes of the $W$ state are distributed among an increasing number of qubits, causing standard tangle-based measures to diminish despite the presence of genuine multipartite entanglement. Therefore, the one-tangle and $\pi$-tangle become inadequate indicators of entanglement for large-$n$ $W$ states.

In the following, we show that the sum of the squared one-tangles can quantify the entanglement of the generalized $W$ state when the number of qubits is large. This measure of entanglement is obtained using Eqs. (7) and (13) as follows:

$$Sum\left(N_{i|\bar{i}}^{2}\left(W\right)\right)=\frac{4(n-1)}{n} \tag{17}$$

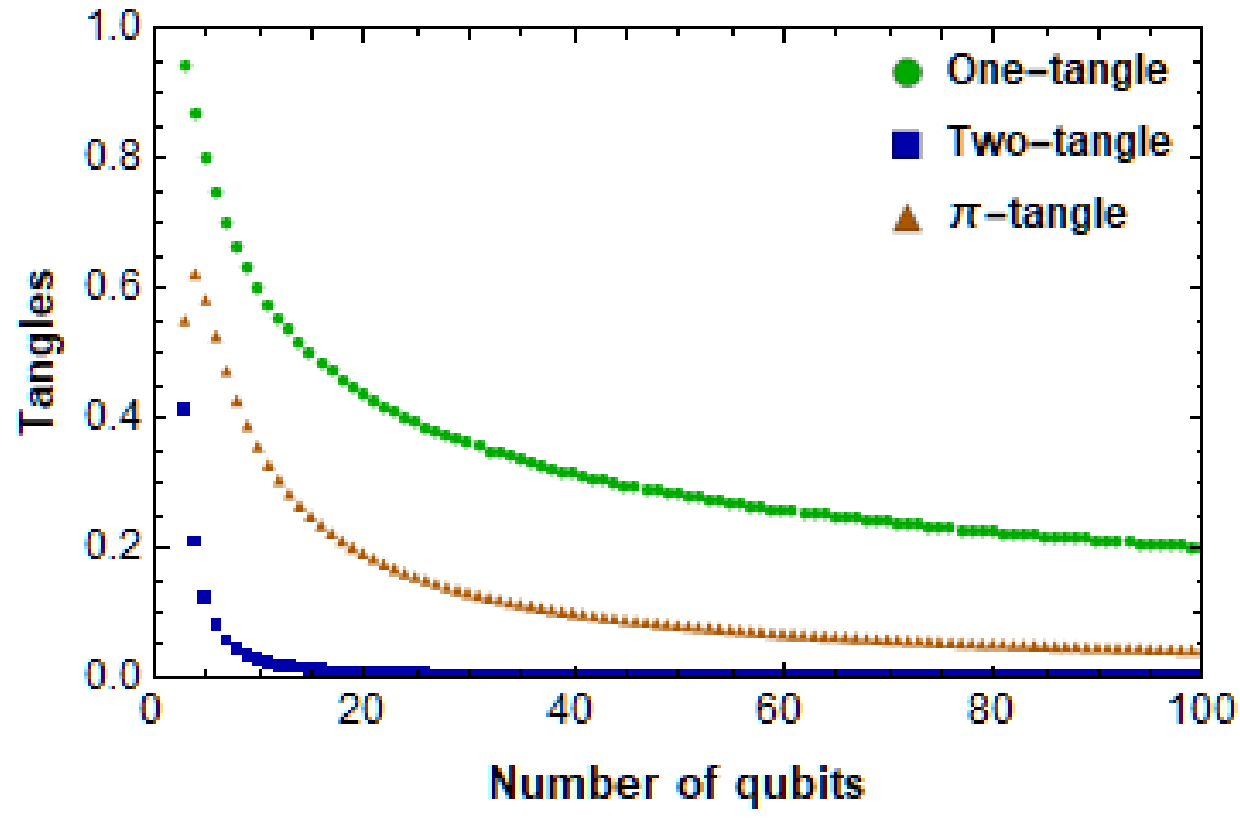

**Fig. 1** The one-tangle (Eq. (13)), two-tangle (Eq. (15)), and $\pi$-tangle (Eq. (16)) of a generalized $W$ state plotted against the number of qubits.

In Fig. 2, the sum of squared one-tangles (Eq. (17)) is plotted as a function of the number of qubits. Unlike conventional measures such as one-tangle or $\pi$-tangle, which decrease toward zero as $n$ increases, the sum of squared one-tangles approaches a finite limit of 4. This occurs because, although the entanglement of individual qubits with the rest decreases as the number of qubits grows, squaring and summing the one-tangles over all qubits captures the total residual entanglement of the system. Physically, this measure reflects the genuinely multipartite correlations present in the generalized $W$ state, effectively quantifying entanglement that persists even in large systems.

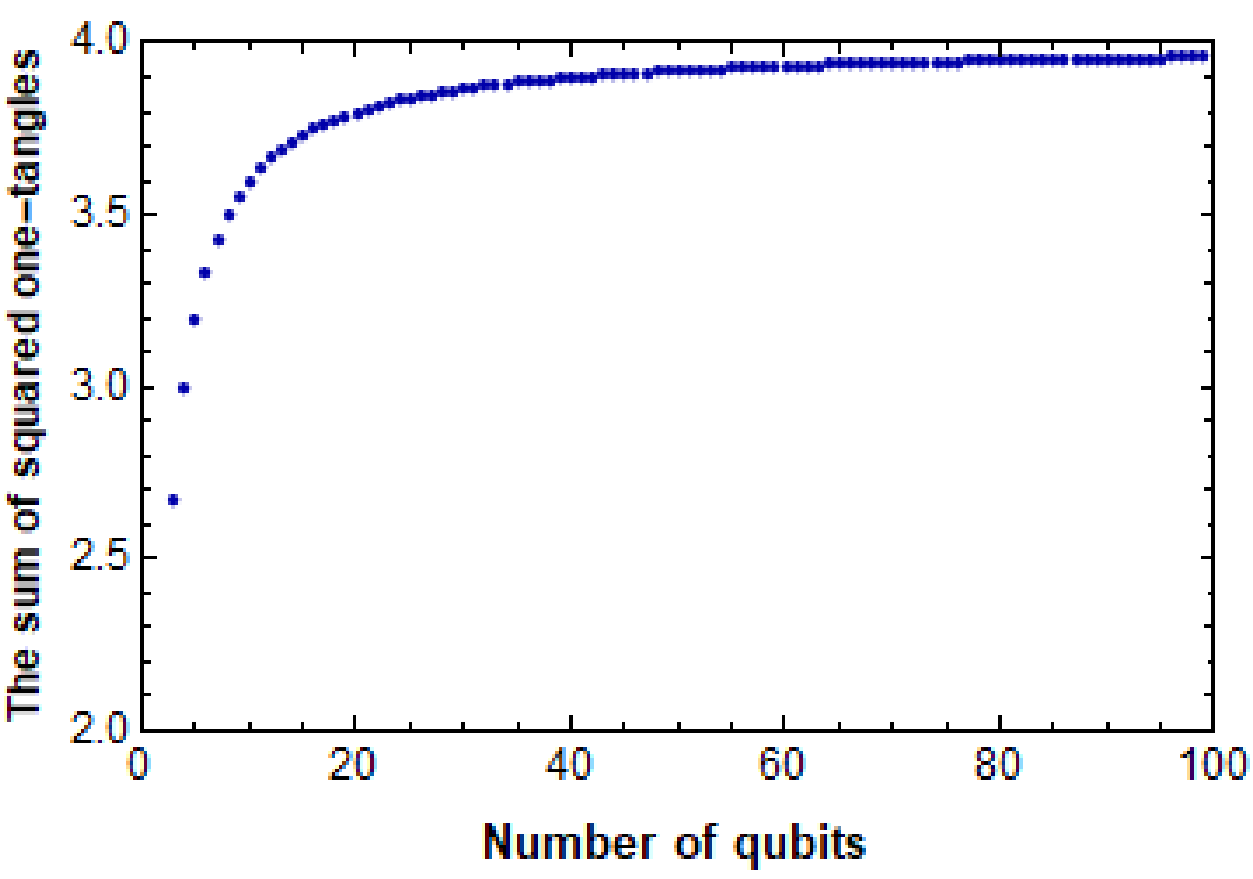

**Fig. 2** The sum of squared one-tangles (Eq. (17)) of the generalized $W$ state plotted against the number of qubits.

## 3.2 The ξ state

The $\xi$ state (as defined in Section 3) can be expressed in terms of the generalized $W$ state as follows:

$$|\xi\rangle = \frac{1}{\sqrt{n+1}}\left(|1\rangle^{\otimes n} + \sqrt{n}\,|W\rangle\right) \tag{18}$$

The density matrix of the $\xi$ state is then given by:

$$\eta_\xi = \frac{1}{n+1}\Big(|1\rangle^{\otimes n}\langle 1|^{\otimes n} + \sqrt{n}\,|W\rangle\langle 1|^{\otimes n} + \sqrt{n}\,|1\rangle^{\otimes n}\langle W| + n\rho_W\Big) \tag{19}$$

where the density matrix of the $W$ state ($\rho_W$) is given in Eq. (8). Next, the $\pi$-tangle is calculated for this state. Given that the $\xi$ state is symmetric, the $\pi$-tangle can be calculated using Eq. (5). Therefore, it is only necessary to determine the $\pi$-tangle with respect to an arbitrary qubit. Here, the first qubit is considered. The partial transpose of the $\xi$ state with respect to the first qubit is as follows:

$$\eta_\xi^{T_1} = \frac{1}{n+1}\Big(\left(|1\rangle^{\otimes n}\langle 1|^{\otimes n} + \sqrt{n}\left(|W\rangle\langle 1|^{\otimes n}\right)\right)^{T_1} + \sqrt{n}\left(|1\rangle^{\otimes n}\langle W|\right)^{T_1} + n\rho_W^{T_1}\Big) \tag{20}$$

where $\rho_W^{T_1}$ is given in Eq. (11). The only non-zero matrix elements are those corresponding to the following bases in the Hilbert space:

$$\begin{aligned}\{&|000...0\rangle, |100...0\rangle, |010...0\rangle, ..., |000...1\rangle,\\ &|110...0\rangle, |101...0\rangle, ..., |100...1\rangle,\\ &|011...1\rangle, |111...1\rangle\}\end{aligned} \tag{21}$$

These bases are the same as those given in relation (10), with the difference that two bases, $|011...1\rangle$ and $|111...1\rangle$, have been added. Therefore, the matrix representation of $\eta_\xi^{T_1}$ in terms of the above bases is as follows:

$$\eta_\xi^{T_1} = \frac{1}{n+1} \times \begin{pmatrix} 0 & 0 & 0 & \cdots & 0 & 1 & 1 & 1 & \cdots & 1 & 0 & 0 \\ 0 & 1 & 0 & \cdots & 0 & 0 & 0 & 0 & \cdots & 0 & 0 & 1 \\ 0 & 0 & 1 & \cdots & 1 & 0 & 0 & 0 & \cdots & 0 & 0 & 0 \\ \vdots & \vdots & \vdots & \ddots & \vdots & 0 & 0 & 0 & \cdots & 0 & \vdots & \vdots \\ 0 & 0 & 1 & \cdots & 1 & 0 & 0 & 0 & \cdots & 0 & 0 & 0 \\ 1 & 0 & 0 & \cdots & 0 & 0 & 0 & 0 & \cdots & 0 & 1 & 0 \\ 1 & 0 & 0 & \cdots & 0 & 0 & 0 & 0 & \cdots & 0 & 1 & 0 \\ 1 & 0 & 0 & \cdots & 0 & 0 & 0 & 0 & \cdots & 0 & 1 & 0 \\ \vdots & \vdots & \vdots & \cdots & \vdots & \vdots & \vdots & \vdots & \cdots & 0 & \vdots & \vdots \\ 1 & 0 & 0 & \cdots & 0 & 0 & 0 & 0 & \cdots & 0 & 1 & 0 \\ 0 & 0 & 0 & \cdots & 0 & 1 & 1 & 1 & \cdots & 1 & 0 & 0 \\ 0 & 1 & 0 & \cdots & 0 & 0 & 0 & 0 & \cdots & 0 & 0 & 1 \end{pmatrix} \tag{22}$$

The only negative eigenvalue of $\eta_\xi^{T_1}$ is $-2\sqrt{(n-1)/2}\,/(n+1)$. Then, using Eq. (4), the one-tangles are calculated as follows:

$$N_{i|\bar{i}}(\xi) = \frac{\sqrt{8(n-1)}}{n+1} \tag{23}$$

To compute the two-tangles, first, the density matrices of the bipartitions are needed. This requires performing the partial trace over $\eta_\xi$ (as defined in Eq. (19)) $n-2$ times. For $n>3$, tracing the first term of $\eta_\xi$ yields $(n+1)^{-1}|11\rangle\langle 11|$, while tracing over the second and third terms eliminates them, and tracing over the fourth term of $\eta_\xi$ results in the density matrix (14). Therefore, the density matrix of the bipartition $(i,k)$ is obtained as follows:

$$\eta_{ik} = \frac{1}{n+1}\begin{pmatrix} n-2 & 0 & 0 & 0 \\ 0 & 1 & 1 & 0 \\ 0 & 1 & 1 & 0 \\ 0 & 0 & 0 & 1 \end{pmatrix}. \tag{24}$$

It is straightforward to demonstrate that $\eta_{ik}$ is separable. In other words, we have:

$$N_{ik}(\xi) = 0. \tag{25}$$

Using Eqs. (5), (23), and (25), the $\pi$-tangle is expressed as follows:

$$\pi_{\xi} = \frac{8(n-1)}{(n+1)^2}. \tag{26}$$

Fig. 3 illustrates the one-tangle (Eq. (23)) and $\pi$-tangle (Eq. (26)) of the $\xi$ state as a function of the number of qubits. Similar to the generalized $W$ state, both the one-tangle and $\pi$-tangle approach zero as the number of qubits increases. In other words, for $n \to \infty$, inequality (3) tends to an equality. This indicates that the one-tangle and $\pi$-tangle are only effective measures of entanglement for the $\xi$ state when the number of qubits is low. In other words, when the number of qubits is large, these measures yield values close to zero for this state.

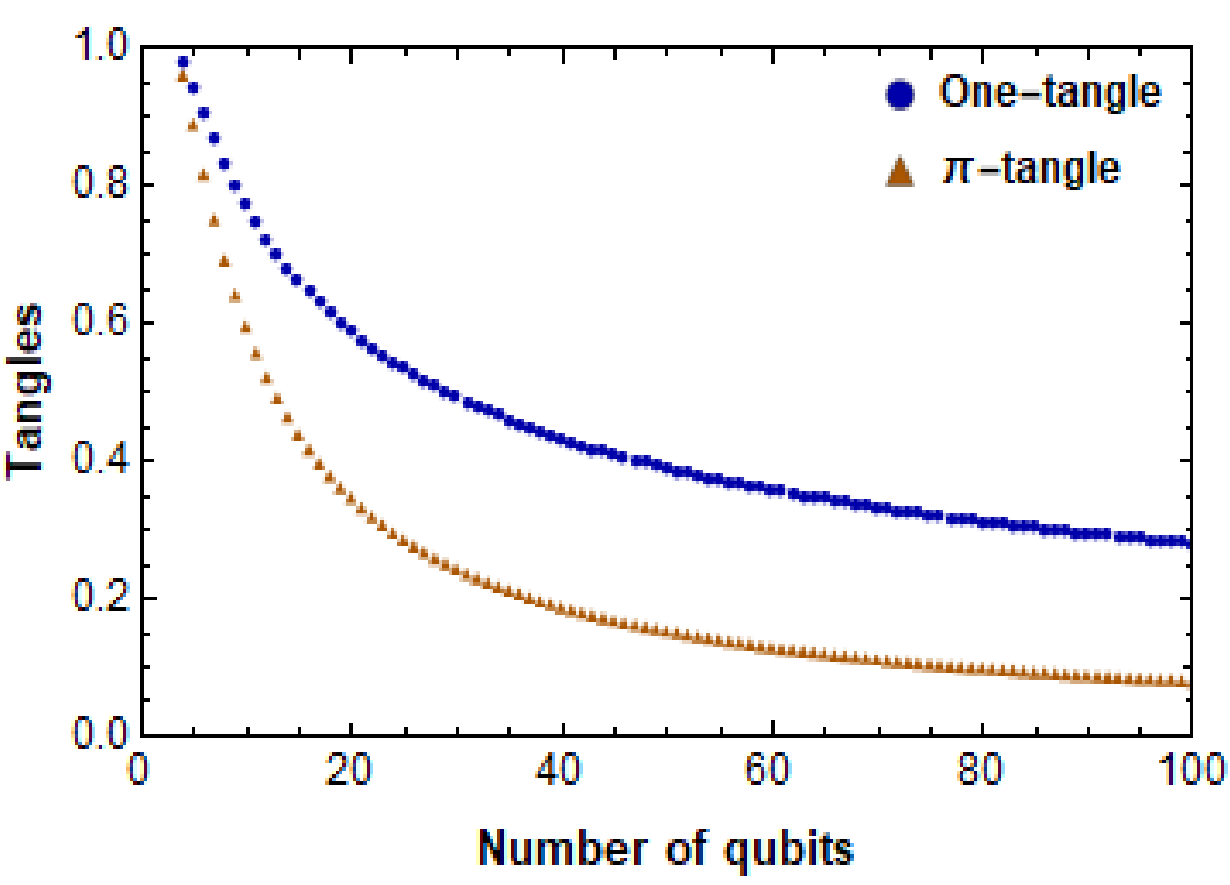

**Fig. 3** The one-tangle (Eq. (23)) and $\pi$-tangle (Eq. (26)) of the $\xi$ state as functions of the number of qubits.

In the following, we show that the sum of squared one-tangles can quantify the entanglement of the $\xi$ state when the number of qubits is large. Using Eqs. (7) and (23), the sum of squared one-tangles is obtained as follows:

$$Sum\left(N_{i|\bar{i}}^2(\xi)\right) = \frac{8n(n-1)}{(n+1)^2}. \tag{27}$$

In Fig. 4, the sum of squared one-tangles (Eq. (27)) is plotted as a function of the number of qubits for the $\xi$ state. Similar to the $W$ state, this sum does not tend toward zero as the number of qubits increases; instead, it approaches a finite nonzero value of 8. This indicates that even as individual one-tangles decrease with increasing qubit number, the total residual entanglement remains significant. The finite asymptotic value reflects the multipartite correlations present in the $\xi$ state and confirms that the sum of squared one-tangles is a reliable measure of entanglement for large-$n$ $\xi$ states, consistent with the behavior observed for the $W$ state.

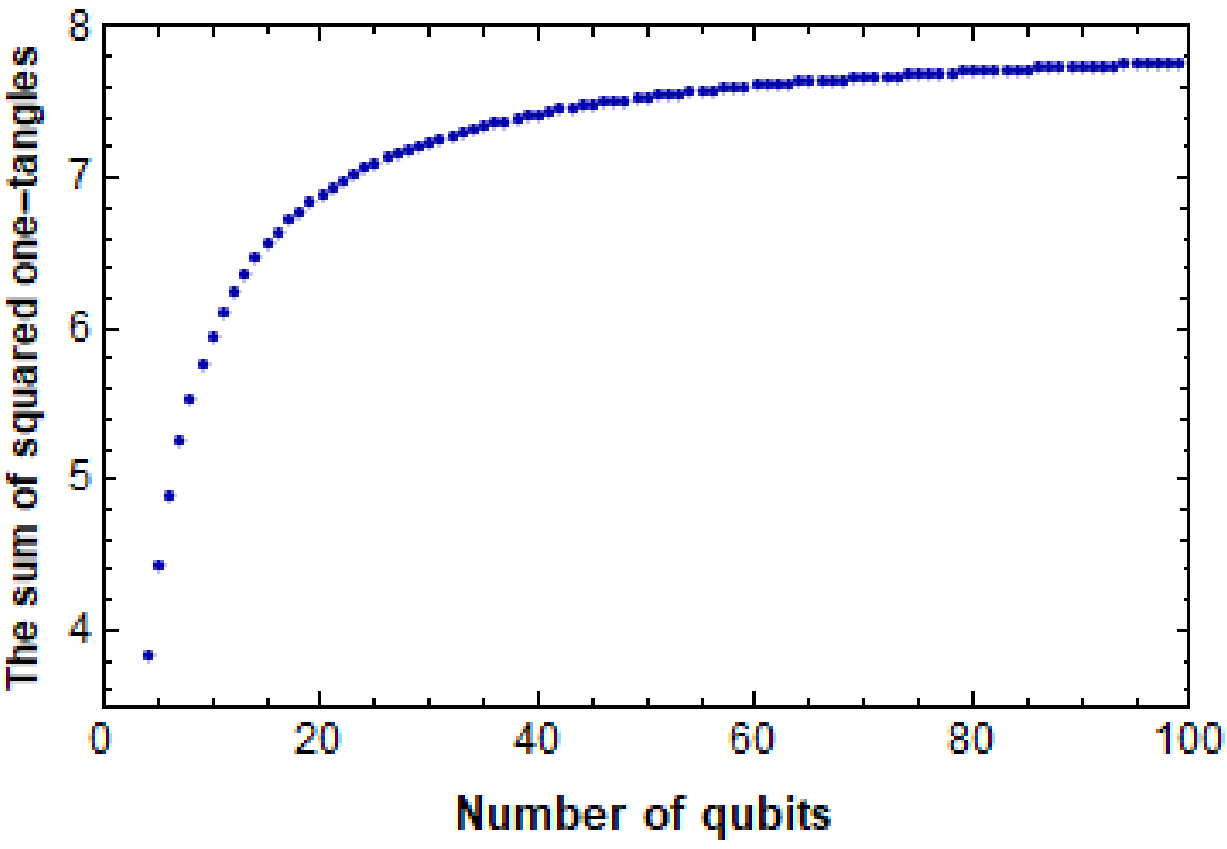

**Fig. 4** The sum of squared one-tangles (Eq. (27)) for the $\xi$ state as a function of the number of qubits.

Since in certain quantum states, like the $W$ and $\xi$ states, increasing the number of qubits causes the monogamy of entanglement to approach equality, we introduce an alternative entanglement inequality that remains strict even as the number of qubits increases. By summing the inequality in Eq. (3) over all qubits, we derive the following relation:

$$\frac{1}{2}\sum_i N_{i|\bar{i}}^2 \geq \sum_{i,k} N_{ik}^2 . \tag{28}$$

The above inequality can be regarded as a strong monogamy inequality that does not approach equality as the number of qubits increases. Moreover, the difference between the two sides of the inequality can be interpreted as the generalized residual entanglement.

$$\Pi = \frac{1}{2}\sum_i N_{i|\bar{i}}^2 - \sum_{i,k} N_{ik}^2. \tag{29}$$

It is straightforward to show that for a pure state, Eq. (29) satisfies the three entanglement measure conditions given in Section 2. Therefore, it is an entanglement measure.

## 4 Conclusion

In this study, we first introduced an entanglement measure based on the tangle, defined as the sum of squared one-tangles. We then proved that this entanglement measure satisfies the three fundamental conditions required for a valid entanglement measure. Next, we considered two $n$-qubit quantum states, the $W$ and $\xi$ states (introduced in Section 3). Since the probability amplitudes of these states depend on the number of qubits, they were selected to examine how increasing the number of qubits affects the entanglement measures discussed in this work. Using the one-tangle, two-tangle, $\pi$-tangle, and the sum of squared one-tangles, we calculated the entanglement of these states. It was observed that as the number of qubits increases, the values of the one-tangle, two-tangle, and $\pi$-tangle all tend to zero. Consequently, the monogamy inequality given in relation (3) converges to equality. This behavior may appear ambiguous, since the system remains entangled even though the conventional measures of entanglement approach zero for large $n$. In contrast, the proposed measure, the sum of squared one-tangles, retains a nonzero value, effectively capturing the residual entanglement in these states. It should be noted that for quantum states whose probability amplitudes are not functions of the number of qubits, such as the GHZ state, the aforementioned entanglement measures do not tend to zero and can still be effectively used to evaluate entanglement in such cases. Therefore, the sum of squared one-tangles is not a suitable measure for this type of states.

Although the entanglement monogamy inequality has been analytically extended to multi-qubit systems and inspired the development of tangle-based measures such as the $\pi$-tangle [24-29], its asymptotic behavior for large systems has often been overlooked. This work explicitly highlights this behavior and proposes approaches to address the limitations it imposes.

Finally, we introduced a stronger form of the entanglement monogamy relation that does not converge to equality as the number of qubits increases. The difference between the two sides of this new inequality can be interpreted as a generalized residual entanglement, which itself serves as an entanglement measure.

**Acknowledgments** This research has received no external funding.

## Author contributions

The author confirms sole responsibility for the conception of the study, theoretical development, calculations, analysis, and manuscript preparation.

**Funding Information** The author received no financial support for this research, authorship, and/or publication of this article.

**Data Availability Statement** No data associated in the manscript.